\documentclass{elsart}

\usepackage{adnd}
\usepackage{longtable}

\usepackage{mathptmx}

\usepackage{amsmath}

\usepackage{amssymb,amstext}
\usepackage[pdftex,
    letterpaper=true,
    hyperindex=true,
    breaklinks=true,
    colorlinks=false,
    citecolor=blue,
    pdftitle={},
    pdfauthor={}]
{hyperref}

\usepackage{graphicx}

\begin{document}

\begin{frontmatter}

\journal{Atomic Data and Nuclear Data Tables}

\copyrightholder{Elsevier Science}

\runtitle{Tungsten}
\runauthor{Fritsch}


\title{Discovery of the Tungsten Isotopes}


\author{A. Fritsch},
\author{J.Q. Ginepro},
\author{M. Heim},
\author{A. Schuh},
\author{A. Shore},
\and
\author{M.~Thoennessen\corauthref{cor}}\corauth[cor]{Corresponding author.}\ead{thoennessen@nscl.msu.edu}

\address{National Superconducting Cyclotron Laboratory and \\ Department of Physics and Astronomy, Michigan State University, \\East Lansing, MI 48824, USA}

\date{2.2.2009} 

\begin{abstract}
Thirty-five tungsten isotopes have so far been observed; the discovery of these isotopes is discussed.  For each isotope a brief summary of the first refereed publication, including the production and identification method, is presented.
\end{abstract}

\end{frontmatter}





\newpage
\tableofcontents
\listofDtables

\vskip5pc

\section{Introduction}\label{s:intro}
The fourth paper in the series of the discovery of isotopes \cite{Gin09,Sch09a,Sch09b}, the discovery of the tungsten isotopes is discussed. Previously, the discovery of cerium \cite{Gin09}, arsenic \cite{Sch09a} and gold \cite{Sch09b} isotopes was discussed. The purpose of this series is to document and summarize the discovery of the isotopes. Guidelines for assigning credit for discovery are (1) clear identification, either through decay-curves and relationships to other known isotopes, particle or $\gamma$-ray spectra, or unique mass and Z-identification, and (2) publication of the discovery in a refereed journal. The authors and year of the first publication, the laboratory where the isotopes were produced as well as the production and identification methods are discussed. When appropriate, references to conference proceedings, internal reports, and theses are included. When a discovery included a half-life measurement the measured value is compared to the currently adapted value taken from the NUBASE evaluation \cite{Aud03} which is based on ENSDF database \cite{ENS08}. In cases where the reported half-life differed significantly from the adapted half-life (up to approximately a factor of two), we searched the subsequent literature for indications that the measurement was erroneous. If that was not the case we credited the authors with the discovery in spite of the inaccurate half-life.

\begin{figure}
 	\centering
	\includegraphics[width=12cm]{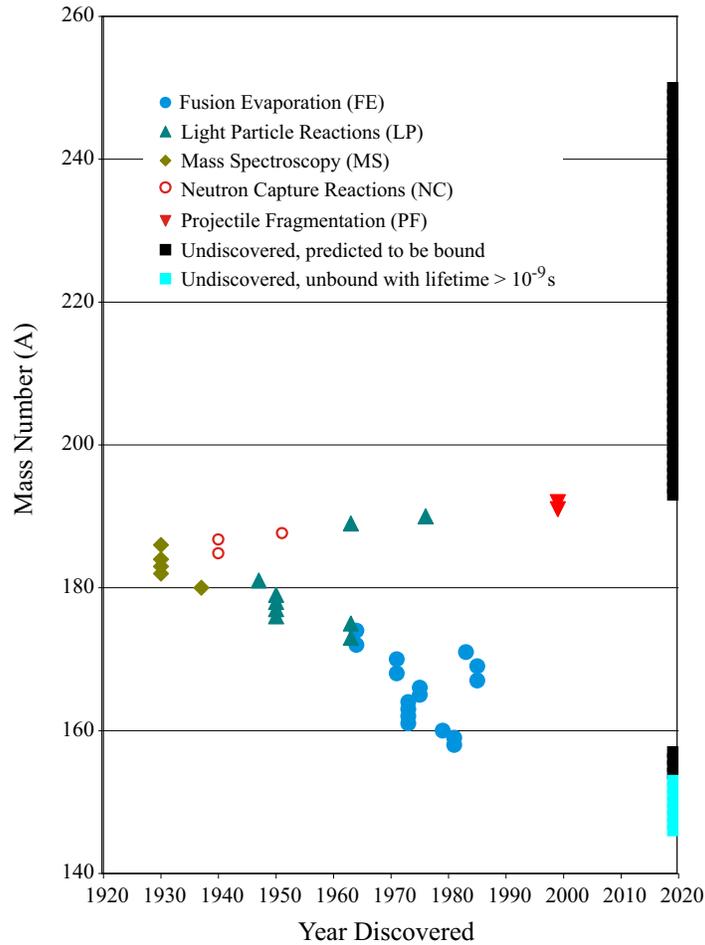}
	\caption{Tungsten isotopes as a function of when they were discovered. The different production methods are indicated. The solid black squares on the right hand side of the plot are isotopes predicted to be bound by the HFB-14 model.  On the proton-rich side the light blue squares correspond to unbound isotopes predicted to have lifetimes larger than $\sim 10^{-9}$~s.}
	\label{f:year}
\end{figure}

\section{Discovery of $^{158-192}$W}
Thirty-five tungsten isotopes from A = $158-192$ have been discovered so far; these include 5 stable, 7 neutron-rich and 23 proton-rich isotopes.  According to the HFB-14 model \cite{Gor07}, tungsten isotopes ranging from $^{154}$W through $^{250}$W should be particle stable. Thus, there remain 62 isotopes to be discovered. In addition, it is estimated that 8 additional nuclei beyond the proton dripline could live long enough to be observed \cite{Tho04}. About one-third of all possible tungsten isotopes have been produced and identified so far.

Figure \ref{f:year} summarizes the year of first discovery for all tungsten isotopes identified by the method of discovery.  The range of isotopes predicted to exist is indicated on the right side of the figure.  Only four different reaction types were used to produce the radioactive tungsten isotopes; heavy-ion fusion evaporation (FE), light-particle reactions (LP), neutron-capture reactions (NC) and projectile fragmentation (PF). The stable isotopes were identified using mass spectroscopy (MS). Heavy ions are all nuclei with an atomic mass larger than A = 4 \cite{Gru77}. Light particles also include neutrons. In the following, the discovery of each tungsten isotope is discussed in detail.

\subsection*{$^{158-159}$W}\vspace{-0.85cm}
In 1981, Hofmann {\it{et al.}} discovered $^{158}$W and $^{159}$W at the Gesellschaft f\"{u}r Schwerionenforschung (GSI) in Darmstadt, Germany, as reported in their paper {\it{New Neutron Deficient Isotopes in the Range of Elements Tm to Pt}} \cite{Hof81}.  Using a 4.4~A$\cdot$MeV nickel beam the isotopes were made in the fusion-evaporation processes $^{106}$Cd($^{58}$Ni,$2p4n$)$^{158}$W and $^{110}$Cd($^{58}$Ni, $2p7n$)$^{159}$W. $^{158}$W was identified by reconstructing its $\alpha$ decay into $^{154}$Hf: ``We explain these observations by the decay chain $^{158}$W $\overset{\alpha}{\rightarrow}$ $^{154}$Hf $\overset{\beta}{\rightarrow}$ $^{154}$Lu $\overset{\beta}{\rightarrow}$ $^{154}$Yb $\overset{\alpha}{\rightarrow}$ $^{150}$Er.'' The half-life of the decay was not measured. $^{159}$W was identified by reconstructing its $\alpha$ decay into $^{155}$Hf of $E_{\alpha}=6299(6)$~keV with a half-life of $t_{1/2}=7.3(27)$~ms: ``Therefore, our observations can easily be described within the frame of the decay chain $^{159}$W $\overset{\alpha}{\rightarrow}$ $^{155}$Hf $\overset{\beta}{\rightarrow}$ $^{155}$Lu $\overset{\alpha}{\rightarrow}$ $^{151}$Tm.'' The half-life agrees with the currently accepted value of 8.2(7)~ms.

\subsection*{$^{160}$W}\vspace{-0.85cm}
Hofmann {\it{et al.}} were the first to discover $^{160}$W in 1979 at GSI as reported in {\it{Alpha Decay Studies of Very Neutron Deficient Isotopes of Hf, Ta, W, and Re}} \cite{Hof79}.  Beams of $^{58}$Ni were incident on various targets of silver, palladium, and rhodium.  Fusion-evaporation products were imbedded into a silicon surface barrier detector after leaving the target. By measuring the products' $\alpha$ decays, rare neutron deficient isotopes, such as $^{160}$W, were found: ``In the investigated reactions the eleven new isotopes ... $^{160}$W ... could be identified.'' The $\alpha$-particle decay energy of $^{160}$W was measured to be 5920(10)~keV. The half-life for this decay was not determined.

\subsection*{$^{161-164}$W}\vspace{-0.85cm}
In 1973, Eastham and Grant were the first to produce the isotopes $^{161}$W, $^{162}$W, $^{163}$W, and $^{164}$W as reported in {\it{Alpha Decay of Neutron-Deficient Isotopes of Tungsten}} \cite{Eas73}. Magnesium beams of energies between 110 and 204~MeV from the Manchester University Hilac were used on samarium targets. $^{161}$W and $^{162}$W were produced in the fusion-evaporation reactions $^{144}$Sm($^{24}$Mg,$xn$) and $^{163,164}$W were both produced in the two reactions $^{144}$Sm($^{24}$Mg,$xn$) and $^{147}$Sm($^{24}$Mg,$xn$). The isotopes were identified by their radioactivity using a helium jet technique. The authors state for the observation of $^{161}$W: ``We make the tentative suggestion that $^{161}$W may decay by emission of an $\alpha$-particle of energy about 5.75~MeV.'' This energy was later confirmed \cite{Hof79}. For $^{162}$W, the $\alpha$ decay energy was found to be $E_{\alpha}=5.53(1)$~MeV and only an upper limit for the lifetime was quoted: ``We have not been able to measure the lifetime of $^{162}$W. Practically no events at all are seen at 5.53 MeV in the observation of a catcher plate flipped out of the helium jet, so the lifetime must be considerably shorter than the dead time of 1/4~s.'' This limit was later found to be incorrect \cite{Hof79}. The $\alpha$-decay energy of $^{163}$W was 5.385(5)~MeV with a half-life of 2.5(3)~s which agrees with the accepted value of 2.8(2)~s. The $\alpha$-decay energy of $^{164}$W was 5.153(5)~MeV with a half-life of 6.3(5)~s. This half-life value is included in the calculation of the currently accepted of 6.3(2) s.

\subsection*{$^{165-166}$W}\vspace{-0.85cm}
$^{165}$W and $^{166}$W were discovered by Toth {\it{et al.}} in 1975 as reported in  {\it{Production and investigation of tungsten $\alpha$ emitters including the new isotopes, $^{165}$W and $^{166}$W}} \cite{Tot75}. The isotopes were produced with $^{16}$O beams from the Oak Ridge isochronous cyclotron bombarding a $^{156}$Dy target. The ORIC gas-jet-capillary system transported the nuclei to a collection chamber where the decay of fusion-evaporation residues was measured. $^{165}$W undergoes $\alpha$ decay with a half-life of 5.1(5)~s and an associated energy of $E_{\alpha}=4.909(5)$~MeV. $^{166}$W decays by $\alpha$ emission with a half-life of 16(3)~s and an associated energy of $E_{\alpha}=4.739(5)$~MeV.  The identification was supported by the following statements: ``... the energies determined in this work for $^{165}$W and $^{166}$W fit well not only as an extension of the data of Eastham and Grant \cite{Eas73} but also into the general $\alpha$-decay systematics in this mass region.''  Furthermore ``... stringent arguments can be presented to exclude the assignment of the new $\alpha$ emitters to isotopes of elements below hafnium.  Thus, ... we believe that the two new $\alpha$ groups represent the $\alpha$ decay of $^{165}$W and $^{166}$W.''
The half-life measurement for $^{165}$W is presently the only measured value and the result for $^{166}$W is included in the accepted value of 19.2(6)~s.

\subsection*{$^{167}$W}\vspace{-0.85cm}
In 1985, Gerl {\it{et al.}} first identified $^{167}$W in their paper {\it{Spectroscopy of $^{166}$W and $^{167}$W and Alignment Effects in Very Neutron-Deficient Tungsten Nuclei}} \cite{Ger85}.  The Australian National University 14UD Pelletron accelerator was used to accelerate a $^{24}$Mg beam. $^{167}$W was created in the fusion reaction $^{147}$Sm($^{24}$Mg,$4n$)$^{167}$W.  The yrast band up to high spins was measured with hyperpure germanium detectors. The statement in the introduction ``The investigation deals with the behaviour of high-spin states in $^{166}$W and $^{167}$W, nuclei which have not previously been studied'' refers to the first observation of high-spin states and the authors were apparently not aware that they were the first to discover $^{167}$W. In turn, in 1989 Meissner {\it{et al.}} was not aware of the Gerl paper in their publication {\it{Decay of the New Isotope $^{167}W$}} \cite{Mei89}.

\subsection*{$^{168}$W}\vspace{-0.85cm}
The observation of $^{168}$W was reported for the first time in 1971 by Stephens {\it{et al.}} as reported in {\it{Some Limitations on the Production of very Neutron-Deficient Nuclei}} \cite{Ste71}. $^{168}$W was produced with a 155 MeV $^{28}$Si beam from the Berkeley Hilac in the fusion-evaporation reaction $^{144}$Sm($^{28}$Si,$2p2n$)$^{168}$W. Yrast $\gamma$ rays were measured up to 8$^+$ with Ge(Li) detectors; ``... the rotational lines of $^{168}$W are very analogous to those of $^{124}$Ba, described above, and their ratio in- and out-of-beam is in excellent agreement with the calculation based on the above value for $k$.'' This paper is referenced as the observation of $^{168}$W only by Dracoulis in 1983 \cite{Dra83}.

\subsection*{$^{169}$W}\vspace{-0.85cm}
In 1985, Recht {\it{et al.}} first observed $^{169}$W as reported in {\it{High-Spin Structure in $^{169}$W and $^{170}$W}} \cite{Rec85}. Neon beams ranging from 105 to 125 ~MeV from the Hahn Meitner Institut Berlin VICKSI accelerator facility bombarded a gadolinium target. The fusion-evaporation reaction $^{154}$Gd($^{20}$Ne,$5n$)$^{169}$W was used to produce $^{169}$W. Gamma-ray spectra were measured with germanium detectors in coincidence with a high-spin filter consisting of NaI scintillation detectors. ``Levels up to about spin 30 in $^{170}$W and up to 57/2 in $^{169}$W have been identified.''

\subsection*{$^{170}$W}\vspace{-0.85cm}
Nadzhakov {\it{et al.}} discovered $^{170}$W in 1971 and presented the results in {\it{New Tungsten Isotopes}} \cite{Nad71}. $^{20,22}$Ne beams accelerated to 145$-$155 MeV by the Dubna U-300 accelerator bombarded isotopically enriched $^{155}$Gd and $^{156}$Gd targets. The isotope was produced in $x$n fusion-evaporation reactions and identified by measuring $\gamma$-ray spectra following chemical separation. The paper states: ``Figure 4 shows the chemical results, while Fig. 5 shows the rise in $^{170}$Ta activity. These results together indicate the presence of the new isotope $^{170}$W with T = 4$\pm$1 min.'' The measured half-life for $^{170}$W is close to the accepted value of 2.42(4)~m.

\subsection*{$^{171}$W}\vspace{-0.85cm}

$^{171}$W was discovered by Arciszewski {\it{et al.}} in 1983 as reported in {\it Band-crossing phenomena in $^{167,168}$Hf and $^{170,171}$W} \cite{Arc83}. The Louvain-la-Neuve CYCLONE Cyclotron accelerated $^{20}$Ne to 110 MeV and $^{171}$W was produced in the fusion-evaporation reaction $^{155}$Gd($^{20}$Ne,4n). Two Compton-suppression spectrometers located at +90$^\circ$ and $-$90$^\circ$ with respect to the beam direction recorded $\gamma$-$\gamma$-coincidences. ``Since the other rotational band could be assigned to $^{169}$W or $^{171}$W, or even to a tantalum isotope (through an xnp channel), excitation function measurements were undertaken... The comparison of fig. 3 clearly shows that the newly observed band belongs to $^{171}$W.'' A previous half-life measurement of $^{171}$W reported a value of 9.0(15)~m \cite{Nad71} which differs by almost a factor of four from the accepted value and was thus not credited for the discovery of $^{171}$W.

\subsection*{$^{172}$W}\vspace{-0.85cm}
$^{172}$W was first reported by Stephens {\it{et al.}} in 1964 in {\it{Properties of High-Spin Rotational States in Nuclei}} \cite{Ste64}. The experimental details were included in a subsequent longer paper \cite{Ste65}. Excitation functions of $^{11}$B, $^{14}$N, and $^{19}$F beams from the Lawrence Radiation Laboratory Hilac on $^{159}$Tb, $^{165}$Ho, and $^{169}$Tm were measured. To produce $^{172}$W, a 117~MeV $^{14}$N beam bombarded a holmium target, resulting in the fusion-evaporation reaction $^{165}$Ho($^{14}$N,7$n$)$^{172}$W. A single wedge-gap electron spectrometer was used to detect conversion electrons. In addition, $\gamma$-ray spectra were recorded with NaI(Tl) and germanium detectors. The original paper only shows the rotational constants as a function of spin. The second paper presents 9 $\gamma$-ray transitions up to spin 18 in $^{172}$W and states:  ``Mass assignments were made on the basis of the change in bombarding energy necessary to go from the maximum of the excitation function of one even nucleus to that of the next lighter one-about 30~MeV per pair of neutrons out. The close similarity in bombarding energy to produce a given reaction in any of these three targets coupled with the results of a number of cases where a given product could be made via more than one reaction leaves no doubt as to the mass assignments.''

\subsection*{$^{173}$W}\vspace{-0.85cm}
In 1963 Santoni {\it{et al.}} reported the discovery of $^{173}$W in {\it{Spectres $\gamma$ et Periodes de Quatre Isotopes de A Impair du Tungstene et Du Tantale}} \cite{San63}. Tantalum oxide was bombarded with protons between 40 and 155 MeV from the Orsay synchrocyclotron. The isotopes were separated using a magnetic separator and their $\gamma$-ray spectra were measured. ``Des cristaux NaI(Tl) 7.5x7.5 cm et 2.5x2.5 cm reli\'es \`a un analyseur \`a 256 canaux ont permis de d\'eterminer les p\'eriodes et d'identifier les spectres $\gamma$ des isotopes 173, 175, 177 et 179 du tungst\`ene et du tantale.'' (7.5x7.5 cm and 2.5x2.5 cm NaI(Tl) crystals with a 256 channel analyzer were used to identify the $\gamma$ spectra of tungsten and tantalum isotopes 173, 175, 177, and 179.) The extracted half-life of $^{173}$W of 16(5)~m is somewhat larger than the accepted value of 7.6(2)~m; however, the measured half-life of the daughter $^{173}$Ta was correct.

\subsection*{$^{174}$W}\vspace{-0.85cm}
$^{174}$W was first reported in the same paper as $^{172}$W by Stephens {\it{et al.}} in 1964 in {\it{Properties of High-Spin Rotational States in Nuclei}} \cite{Ste64}. The experimental details were included in the subsequent longer paper \cite{Ste65}. Excitation functions of $^{11}$B, $^{14}$N, and $^{19}$F beams from the Lawrence Radiation Laboratory Hilac on $^{159}$Tb, $^{165}$Ho, and $^{169}$Tm were measured. $^{174}$W was produced with a 83~MeV $^{11}$B beam bombarding a thulium target in the fusion-evaporation reaction $^{169}$Tm($^{11}$B,6$n$)$^{174}$W. A single wedge-gap electron spectrometer was used to detect conversion electrons. In addition, $\gamma$-ray spectra were recorded with NaI(Tl) and germanium detectors. The original paper only shows the rotational constants as a function of spin. The second paper presents 7 $\gamma$-ray transitions up to spin 14 in $^{174}$W. The half-life of $^{174}$W was first reported one month prior to the submission of the first Stephens paper in an internal report by Santoni and Valentin \cite{San64} and independently a year later in a refereed publication by Demeter {\it{et al.}} \cite{Dem65}.

\subsection*{$^{175}$W}\vspace{-0.85cm}
In 1963 Santoni {\it{et al.}} reported the discovery of $^{175}$W together with the discovery of $^{173}$W in {\it{Spectres $\gamma$ et Periodes de Quatre Isotopes de A Impair du Tungstene et Du Tantale}} \cite{San63}. Tantalum oxide was bombarded with protons between 40 and 155 MeV from the Orsay synchrocyclotron. The isotopes were separated using a magnetic separator and their $\gamma$-ray spectra were measured. ``Des cristaux NaI(Tl) 7.5x7.5 cm et 2.5x2.5 cm reli\'es \`a un analyseur \`a 256 canaux ont permis de d\'eterminer les p\'eriodes et d'identifier les spectres $\gamma$ des isotopes 173, 175, 177 et 179 du tungst\`ene et du tantale.'' (7.5x7.5 cm and 2.5x2.5 cm NaI(Tl) crystals with a 256 channel analyzer were used to identify the $\gamma$ spectra of tungsten and tantalum isotopes 173, 175, 177, and 179.) The half-life of $^{175}$W was determined to be 34(1) m. This value is consistent with the presently accepted value of 35.2(6) m.

\subsection*{$^{176-179}$W}\vspace{-0.85cm}
The isotopes $^{176}$W, $^{177}$W, $^{178}$W, and $^{179}$W were discovered by Wilkinson from Berkeley in 1950 as reported in  {\it{Neutron Deficient Radioactive Isotopes of Tantalum and Wolfram}} \cite{Wil50}.  Protons from the 184-cyclotron directed on a tantalum target created the isotopes. ``The bombardment of tantalum with protons of energy 10 to 70 Mev has led to the characterization of five new radioactive isotopes of wolfram.'' Wilkinson counted the observation of an isomeric state in $^{179}$W as the fifth isotope. They were identified following chemical separation by the measurement of K X-rays, electrons and $\gamma$ radiation. The half-life of $^{176}$W was measured to be 80(5)~m which is near the accepted value of 2.5(1)~h. The extracted values for $^{177}$W (130(3)~m) and $^{178}$W (21.5(1)~d) are included in the accepted values of 132(2)~m and 21.6(3)~d, respectively. The half-life for $^{179}$W of 30(1)~m is close to the accepted value of 37.05(16)~m. It should be mentioned that Wilkinson and Hicks had reported a 135~m half-life in 1948; however, they could not uniquely assign it to a specific tungsten isotope (either $^{178}$W or $^{179}$W) \cite{Wil48}.

\subsection*{$^{180}$W}\vspace{-0.85cm}
Dempster reported the existence of the stable isotope $^{180}$W in 1937 in {\it{The Isotopic Constitution of Tungsten}} \cite{Dem37}.  Although there was prior evidence for its existence, impurities of the tungsten electrodes prevented a firm identification. ``With pure tungsten electrodes, six photographs have been made showing the isotope at 180, and by varying the time of exposure, its intensity was estimated as approximately one one-hundredth of that of the isotope at 183.  In the earlier photographs, the faint isotope was also found on two photographs of doubly charged ions, on two of triply charged ions, and on one of quadruply charged ions.  Thus there can be no doubt that tungsten has a fifth faint stable isotope at mass 180.''

\subsection*{$^{181}$W}\vspace{-0.85cm}
In 1947, Wilkinson identified $^{181}$W at the University of California, Berkeley, as reported in {\it{A New Isotope of Tungsten}} \cite{Wil47}. By bombarding 20~MeV deuterons from the Crocker Laboratory cyclotron on a thin tantalum foil $^{181}$W was produced in the reaction $^{181}$Ta($d$, $2n$)$^{181}$W.  Electrons, X-rays, and $\gamma$ rays were measured. ``The tungsten fraction contained a previously unreported, single radioactivity of half-life 140$\pm$2 days.'' This value is close to the currently accepted value of 121.2(2)~d.

\subsection*{$^{182-184}$W}\vspace{-0.85cm}
Aston identified the stable tungsten isotopes $^{182}$W, $^{183}$W and$^{184}$W in 1930 at the Cavendish Laboratory in Cambridge as reported in {\it{Constitution of Tungsten}} \cite{Ast30}.  The observation was ``made possible by the preparation of the volatile carbonyl, W(CO)$_6$, by Dr. A. v. Grosse, of Berlin. It was to be expected from the greater atomic weight that the photographic effect would be feeble, and only by means of very sensitive plates were lines of satisfactory intensity obtained.''

\subsection*{$^{185}$W}\vspace{-0.85cm}
In 1940, Minawaka reported the discovery of $^{185}$W in {\it{Neutron-Induced Radioactivity of Tungsten}} \cite{Min40}. Pure metallic tungsten powder was irradiated with slow and fast neutrons produced in nuclear reactions in the Tokyo cyclotron. A Lauritsen type electroscope, chemical separation and measurements with a thin-walled G-M counter and a magnetic field were used to identify $^{185}$W. ``From the relative intensities of both periods ..., it was found that the shorter period is produced practically only with slow neutrons and the longer one both with fast and slow neutrons.  The above results lead to the conclusion that ... 77-day activity [is due] to W$^{185}$.'' This half-life (77(3)~d) agrees with the accepted value of 75.1(3)~d. It should be mentioned that Fajans and Sullivan confirmed the result by Minawaka later in the same year \cite{Faj40}.

\subsection*{$^{186}$W}\vspace{-0.85cm}
Aston identified the stable tungsten isotope $^{186}$W together with the isotopes $^{182-184}$W in 1930 at the Cavendish Laboratory in Cambridge as reported in {\it{Constitution of Tungsten}} \cite{Ast30}. ``Tungsten proves to have four isotopes, of which the strongest two give lines of practically identical intensity.''

\subsection*{$^{187}$W}\vspace{-0.85cm}
In 1940, Minawaka reported the discovery of $^{187}$W together with the discovery of $^{185}$W in {\it{Neutron-Induced Radioactivity of Tungsten}} \cite{Min40}. Pure metallic tungsten powder was irradiated with slow and fast neutrons produced in nuclear reactions in the Tokyo cyclotron. A Lauritsen type electroscope, chemical separation and measurements with a thin-walled G-M counter and a magnetic field were used to identify $^{187}$W. ``From the relative intensities of both periods ..., it was found that the shorter period is produced practically only with slow neutrons and the longer one both with fast and slow neutrons.  The above results lead to the conclusion that 24-hour activity is due to W$^{187}$ ...'' This half-life (24.0(1)~h) is consistent with the accepted value of 23.72(6)~h. Previously, an $\sim$ 24~h half-life had been measured but could not be assigned to a specific tungsten isotope \cite{Fer35,McL35,Jae37}. It should be mentioned that Fajans and Sullivan confirmed the result by Minawaka later in the same year \cite{Faj40}.

\subsection*{$^{188}$W}\vspace{-0.85cm}
Lindner and Coleman reported the discovery of $^{188}$W in 1951 in {\it{The Identification of W$^{188}$ Formed in Neutron-Activated Tungsten by a Chemical Separation of Re$^{188}$}} \cite{Lin51a}. $^{188}$W was formed through successive neutron capture from $^{186}$W by neutron irradiation in a nuclear reactor.  ``A new radioisotope of tungsten, mass 188, which was formed by successive neutron capture by the heaviest stable tungsten isotope, W$^{186}$, has been indirectly established in the presence of very large levels of other radio-tungsten isotopes. This was accomplished by observing the activity of the known Re$^{188}$ which arises as a result of the decay of the W$^{188}$.'' The measured half-life of 65(5)~d agrees with the accepted value of 69.78(5)~d. Lindner also published the results in a separate paper later in the same year \cite{Lin51b}.

\subsection*{$^{189}$W}\vspace{-0.85cm}
In 1963 Flegenheimer {\it{et al.}} from the Comisi\'{o}n Nacional de Energ\'{i}a At\'{o}mica, Buenos Aires, Argentina, discovered $^{189}$W in {\it{The $^{189}$W - $^{189}$Re Decay Chain}} \cite{Fle63}. Fast neutrons were produced by the bombardment of beryllium with 28 MeV deuterons in the synchrocyclotron. $^{189}$W was produced via the ($n$,$\alpha$) reaction on an osmium target. The half-life was measured following chemical separation. ``No 11-minutes tungsten nuclide was found after a W($d$,$p$) reaction, which excludes mass numbers 185 and 187.  We therefore assign the mass number 189 to this half-life.'' The half-life agrees with the present value of 11.6(3)~m.

\subsection*{$^{190}$W}\vspace{-0.85cm}
In 1976 Haustein {\it{et al.}} observed $^{190}$W for the first time and they reported it in {\it{New neutron-rich isotope: $^{190}$W}} \cite{Hau76}. ``A new neutron-rich isotope, $^{190}$W, was produced by $^{192}$Os($n$,2$pn$)$^{190}$W (25-200~MeV neutrons) and by $^{192}$Os($p$,3$p$)$^{190}$W (92~MeV protons).'' The protons were accelerated by the Brookhaven linac injector of the alternating gradient synchrotron and the neutron irradiations were performed in the MEIN facility. The half-life was shown to be 30.0(15)~m, which is still the only measured half-life of $^{190}$W.

\subsection*{$^{191-192}$W}\vspace{-0.85cm}
In 1999, Benlliure {\it{et al.}} created $^{191}$W and $^{192}$W isotope as reported in {\it{Production of neutron-rich isotopes by cold fragmentation in the reaction $^{197}$Au + Be at 950~$A$~MeV}} \cite{Ben99}.  A 950~A$\cdot$MeV $^{197}$Au beam from the SIS synchrotron of GSI was incident on a beryllium target, resulting in projectile fragmentation.  The FRS fragment separator was used to select isotopes with a specific mass-to-charge ratio.  ``The mass resolution achieved in this measurement was $A/\Delta A\approx400$ ... the isotopes ... $^{191}$W, $^{192}$W ... were clearly identified for the first time. Only isotopes with a yield higher than 15 counts were considered as unambiguously identified.''

\section{Summary}
It is interesting that six tungsten isotopes ($^{167-169}$W, $^{171-172}$W and $^{174}$W) were first identified in high-spin $\gamma$-ray spectroscopy experiments where the authors were not aware of the fact that they were the first to observe the specific isotope. The first measurement of the half-life of $^{171}$W was significantly different from the later accepted values and was not credited with the discovery. The half-lives of two other isotopes ($^{177}$W and $^{187}$W) were observed first without being assigned to the specific isotopes.

\ack

This work was supported by the National Science Foundation under grants No. PHY06-06007 (NSCL) and PHY07-54541 (REU). MH was supported by NSF grant PHY05-55445. JQG acknowledges the support of the Professorial Assistantship Program of the Honors College at Michigan State University.


\begin{thebibliography}{00}

\bibitem{Gin09} J.Q. Ginepro, J. Snyder, and M. Thoennessen, At. Data Nucl. Data Tables, in print (2009)
\bibitem{Sch09a} A. Shore, A. Fritsch, M. Heim, A. Schuh, and M. Thoennessen, submitted to At. Data Nucl. Data Tables (2009)
\bibitem{Sch09b} A. Schuh, A. Fritsch, G.Q. Ginepro, M. Heim, A. Shore, and M. Thoennessen, submitted to At. Data Nucl. Data Tables (2009)
\bibitem{Aud03} G. Audi, O. Bersillon, J. Blachot, and A.H. Wapstra, Nucl. Phys. A {\bf 729}, 3 (2003)
\bibitem{ENS08} ENSDF, Evaluated Nuclear Structure Data File, mainted by the National Nuclear Data Center at Brookhaven National Laboratory, published in Nuclear Data Sheets (Academic Press, Elsevier Science).
\bibitem{Gor07} S. Goriely, M. Samyn, and J.M. Pearson, Phys. Rev. C {\bf 75}, 64312 (2007)
\bibitem{Tho04} M. Thoennessen, Rep. Prog. Phys. {\bf 67}, 1187 (2004)
\bibitem{Gru77} H.A. Grunder and F.B. Selph, Annu. Rev, Nucl. Sci., {\bf27}, 353 (1977)
\bibitem{Hof81} S. Hofmann, G. M{\"{u}}ntzenberg, F. He{\ss}berger, W. Reisdorf, and P. Armbruster, Z. Phys. A {\bf 299}, 281 (1981)
\bibitem{Hof79} S. Hofmann, W. Faust, G. M{\"{u}}ntzenberg, W. Reisdorf, and P. Armbruster, Z. Phys. A {\bf 291}, 53 (1979)
\bibitem{Eas73} D.A. Eastham and I.S. Grant, Nucl. Phys. A {\bf 208}, 119 (1973)
\bibitem{Tot75} K.S. Toth, W.D. Schmidt-Ott, C.R. Bingham, and M.A. Ijaz, Phys. Rev. C {\bf 12}, 533 (1975)
\bibitem{Ger85} J. Gerl, G.D. Dracoulis, A.P. Byrne, A.R. Poletti, S.J. Poletti, and A.E. Stuchbery, Nucl. Phys. A {\bf 443}, 348 (1985)
\bibitem{Mei89} F. Meissner, W. D. Schmidt-Ott, V. Freystein, T. Hild, E. Runte, H. Salewski, and R. Michaelsen, Z. Phys. A {\bf 332}, 153 (1989)
\bibitem{Ste71} F.S. Stephens, J.R. Leigh, and R.M. Diamond, Nucl. Phys. A {\bf 170}, 321 (1971)
\bibitem{Dra83} G.D. Dracoulis, H. H{\"{u}}bel, A.P. Byrne, and R.F. Davie, Nucl. Phys. A {\bf 405}, 363 (1983)
\bibitem{Rec85} J. Recht, Y.K. Agarwal, K.P. Blume, M. Guttormsen, and H. H{\"{u}}bel, Nucl. Phys. A {\bf 440}, 366 (1985)
\bibitem{Arc83} H.F.R. Arciszewski, H.J.M Aarts, R. Kamermans, C.J. Van Der Poel, R. Holzmann, M.-A. Van Hove, J. Vervier, M. Huyse, G. Lhersonneau, R.V.F. Janssens, and M.J.A. De Voigt, Nucl. Phys. A {\bf 401}, 531 1983)
\bibitem{Nad71} E. Nadzhakov, B. Bochev, T. Venkova, V. Mikhailova, M. Mikhailov, T. Kutsarova, G. Radonov, and R. Kalpakchieva, Izv. Akad. Nauk SSSR, Ser. Fiz. {\bf 35}, 2207 (1971), Bull. Acad. Sci. USSR, Phys. Ser. {\bf 35}, 2004 (1972)
\bibitem{Ste64} F.S. Stephens, N. Lark, and R.M. Diamond, Phys. Rev. Lett. {\bf 12}, 225 (1964)
\bibitem{Ste65} F.S. Stephens, N.L. Lark, and R.M. Diamond, Nucl. Phys. {\bf 63}, 82 (1965)
\bibitem{San63} A. Santoni, A. Caruette, and J. Valentin, J. Phys. France {\bf 24}, 407 (1963)
\bibitem{San64} A. Santoni and J. Valentin, Phys. Nucl. Annuaire 1962-1963, Faculte Sci.L'Univ. Paris Inst. Rad., p. 41 (January 1964)
\bibitem{Dem65} I. Demeter, K.H. Sil, E. Nadjakov, and N.G. Zaitseva, Phys. Lett. {\bf 19}, 47 (1965)
\bibitem{Wil50} G. Wilkinson, Phys. Rev. {\bf 80}, 495 (1950)
\bibitem{Wil48} G. Wilkinson and H. Hicks, Phys. Rev. {\bf 74}, 1733 (1948)
\bibitem{Dem37} A.J. Dempster, Phys. Rev. {\bf 52}, 1074 (1937)
\bibitem{Wil47} G. Wilkinson, Phys. Rev. Lett. {\bf 160}, 864 (1947)
\bibitem{Ast30} F.W. Aston, Nature {\bf 126}, 913 (1930)
\bibitem{Min40} O. Minakawa, Phys. Rev. {\bf 57}, 1189 (1940)
\bibitem{Faj40} K. Fajans and W.H. Sullivan, Phys. Rev. {\bf 58}, 276 (1940)
\bibitem{Fer35} E. Fermi, E. Amaldi, O. d'Agostino, F. Rasetti, and E. Segre, Proc. Roy. Soc. {\bf 149}, 522 (1935)
\bibitem{McL35} J.C. McLennan, L.G. Grimmett and J. Read, Nature {\bf 135}, 147 (1935)
\bibitem{Jae37} R. Jaeckel, Zeits. f. Physik {\bf 104}, 762 (1937)
\bibitem{Lin51a} M. Lindner and J.S. Coleman, J. Am. Chem. Soc. {\bf 73}, 1610 (1951)
\bibitem{Lin51b} M. Lindner, Phys. Rev. {\bf 84}, 240 (1951)
\bibitem{Fle63} J. Flegenheimer, G. B. Bar{\'{o}}, and M. Viirsoo, Radiochim. Acta {\bf 2}, 7 (1963)
\bibitem{Hau76} P.E. Haustein, E.M. Franz, S. Katcoff, and N.A. Morcos, Phys. Rev. C {\bf 14}, 645 (1976)
\bibitem{Ben99} J. Benlliure, K.H. Schmidt, D. Cortina-Gil, T. Enqvist, F. Farget, A. Heinz, A.R. Junghas, J. Pereira, and J. Taieb, Nucl. Phys. A {\bf 660}, 87 (1999)

\end{thebibliography}

\newpage

\section*{EXPLANATION OF TABLE}\label{sec.eot}
\addcontentsline{toc}{section}{EXPLANATION OF TABLE}

\renewcommand{\arraystretch}{1.0}

\begin{tabular*}{0.95\textwidth}{@{}@{\extracolsep{\fill}}lp{5.5in}@{}}
\textbf{TABLE I.}
	& \textbf{Discovery of Tungsten Isotopes }\\
\\

Isotope & Tungsten isotope \\
First Author & First author of refereed publication \\
Journal & Journal of publication \\
Ref. & Reference \\
Method & Production method used in the discovery: \\
  & FE: fusion evaporation \\
  & LP: light-particle reactions (including neutrons) \\
  & MS: mass spectroscopy \\
  & NC: neutron-capture reactions \\
  & PF: projectile fragmentation or projectile fission \\
Laboratory & Laboratory where the experiment was performed\\
Country & Country of laboratory\\
Year & Year of discovery \\
\end{tabular*}
\label{tableI}

\newpage
\datatables

\setlength{\LTleft}{0pt}
\setlength{\LTright}{0pt}


\setlength{\tabcolsep}{0.5\tabcolsep}

\renewcommand{\arraystretch}{1.0}


\begin{longtable}[c]{%
@{}@{\extracolsep{\fill}}r@{\hspace{5\tabcolsep}} llllllll@{}}
\caption[Discovery of Tungsten Isotopes]%
{Discovery of Tungsten isotopes}\\[0pt]
\caption*{\small{See page \pageref{tableI} for Explanation of Tables}}\\
\hline
\\[100pt]
\multicolumn{8}{c}{\textit{This space intentionally left blank}}\\
\endfirsthead
Isotope & First Author & Journal & Ref. & Method & Laboratory & Country & Year \\

$^{158}$W & S. Hofmann & Z. Phys. A & Hof81 & FE & Darmstadt & Germany &1981 \\
$^{159}$W & S. Hofmann & Z. Phys. A & Hof81 & FE & Darmstadt & Germany &1981 \\
$^{160}$W & S. Hofmann & Z. Phys. A & Hof79 & FE & Darmstadt & Germany &1979 \\
$^{161}$W & D.A. Eastham & Nucl. Phys. A & Eas73 & FE & Manchester & UK &1973 \\
$^{162}$W & D.A. Eastham & Nucl. Phys. A & Eas73 & FE & Manchester & UK &1973 \\
$^{163}$W & D.A. Eastham & Nucl. Phys. A & Eas73 & FE & Manchester & UK &1973 \\
$^{164}$W & D.A. Eastham & Nucl. Phys. A & Eas73 & FE & Manchester & UK &1973 \\
$^{165}$W & K.S. Toth & Phys. Rev. C & Tot75 & FE & Oak Ridge & USA &1975 \\
$^{166}$W & K.S. Toth & Phys. Rev. C & Tot75 & FE & Oak Ridge & USA &1975 \\
$^{167}$W & J. Gerl & Nucl. Phys. A & Ger85 & FE & Canberra & Australia &1985 \\
$^{168}$W & F.S. Stephens & Nucl. Phys. A & Ste71 & FE & Berkeley & USA &1971 \\
$^{169}$W & J. Recht & Nucl. Phys. A & Rec85 & FE & Berlin & Germany &1985 \\
$^{170}$W & E. Nadzhakov & Izv. Akad. Nauk. & Nad71 & FE & Dubna & Russia &1971 \\
$^{171}$W & H.F.R. Arciszewski & Nucl. Phys. A & Arc83 & FE & Louvain & Belgium &1983 \\
$^{171}$W & A. Szymanski & Radiochim. Acta & Szy87 & FE & Harwell & UK &1987 \\
$^{172}$W & F.S. Stephens & Phys. Rev. Lett. & Ste64 & FE & Berkeley & USA &1964 \\
$^{173}$W & A. Santoni & J. Phys. (France) & San63 & LP & Orsay & France &1963 \\
$^{174}$W & F.S. Stephens & Phys. Rev. Lett. & Ste64 & FE & Berkeley & USA &1964 \\
$^{175}$W & A. Santoni & J. Phys. (France) & San63 & LP & Orsay & France &1963 \\
$^{176}$W & G. Wilkinson & Phys. Rev. & Wil50 & LP & Berkeley & USA &1950 \\
$^{177}$W & G. Wilkinson & Phys. Rev. & Wil50 & LP & Berkeley & USA &1950 \\
$^{178}$W & G. Wilkinson & Phys. Rev. & Wil50 & LP & Berkeley & USA &1950 \\
$^{179}$W & G. Wilkinson & Phys. Rev. & Wil50 & LP & Berkeley & USA &1950 \\
$^{180}$W & A.J. Dempster & Phys. Rev. & Dem37 & MS & Chicago & USA &1937 \\
$^{181}$W & G. Wilkinson & Phys. Rev. Lett. & Wil47 & LP & Berkeley & USA &1947 \\
$^{182}$W & F.W. Aston & Nature & Ast30 & MS & Cambridge & UK &1930 \\
$^{183}$W & F.W. Aston & Nature & Ast30 & MS & Cambridge & UK &1930 \\
$^{184}$W & F.W. Aston & Nature & Ast30 & MS & Cambridge & UK &1930 \\
$^{185}$W & O. Minakawa & Phys. Rev. & Min40 & NC & Tokyo & Japan &1940 \\
$^{186}$W & F.W. Aston & Nature & Ast30 & MS & Cambridge & UK &1930 \\
$^{187}$W & O. Minakawa & Phys. Rev. & Min40 & NC & Tokyo & Japan &1940 \\
$^{188}$W & M. Lindner & J. Am. Chem. Soc. & Lin51 & NC & Washington State & USA &1951 \\
$^{189}$W & J. Flegenheimer & Radiochim. Acta & Fle63 & LP & Buenos Aires & Argentina &1963 \\
$^{190}$W & P.E. Haustein & Phys. Rev. C & Hau76 & LP & Brookhaven & USA &1976 \\
$^{191}$W & J. Benlliure & Nucl. Phys. A & Ben99 & PF & Darmstadt & Germany &1999 \\
$^{192}$W & J. Benlliure & Nucl. Phys. A & Ben99 & PF & Darmstadt & Germany &1999 \\

\end{longtable}
\newpage


\normalsize

\begin{theDTbibliography}{1956He83}
\bibitem[Arc83]{Arc83t} H.F.R. Arciszewski, H.J.M Aarts, R. Kamermans, C.J. Van Der Poel, R. Holzmann, M.-A. Van Hove, J. Vervier, M. Huyse, G. Lhersonneau, R.V.F. Janssens, and M.J.A. De Voigt, Nucl. Phys. A {\bf 401}, 531 1983)
\bibitem[Ast30]{Ast30t} F.W. Aston, Nature {\bf 126}, 913 (1930)
\bibitem[Ben99]{Ben99t} J. Benlliure, K.H. Schmidt, D. Cortina-Gil, T. Enqvist, F. Farget, A. Heinz, A.R. Junghans, J. Pereira, and J. Taieb, Nucl. Phys. A {\bf 660}, 87 (1999)
\bibitem[Dem37]{Dem37t} A.J. Dempster, Phys. Rev. {\bf 52}, 1074 (1937)
\bibitem[Eas73]{Eas73t} D.A. Eastham and I.S. Grant, Nucl. Phys. A {\bf 208}, 119 (1973)
\bibitem[Fle63]{Fle63t} J. Flegenheimer, G. B. Bar{\'{o}}, and M. Viirsoo, Radiochim. Acta {\bf 2}, 7 (1963)
\bibitem[Ger85]{Ger85t} J. Gerl, G.D. Dracoulis, A.P. Byrne, A.R. Poletti, S.J. Poletti, and A.E. Stuchbery, Nucl. Phys. A {\bf 443}, 348 (1985)
\bibitem[Hau76]{Hau76t} P.E. Haustein, E.M. Franz, S. Katcoff, and N.A. Morcos, Phys. Rev. C {\bf 14}, 645 (1976)
\bibitem[Hof79]{Hof79t} S. Hofmann, W. Faust, G. M{\"{u}}ntzenberg, W. Reisdorf, and P. Armbruster, Z. Phys. A {\bf 291}, 53 (1979)
\bibitem[Hof81]{Hof81t} S. Hofmann, G. M{\"{u}}ntzenberg, F. He{\ss}berger, W. Reisdorf, and P. Armbruster, Z. Phys. A {\bf 299}, 281 (1981)
\bibitem[Lin51]{Lin51t} M. Lindner and J.S. Coleman, J. Am. Chem. Soc. {\bf 73}, 1610 (1951)
\bibitem[Min40]{Min40t} O. Minakawa, Phys. Rev. {\bf 57}, 1189 (1940)
\bibitem[Nad71]{Nad71t} E. Nadzhakov, B. Bochev, T. Venkova, V. Mikhailova, M. Mikhailov, T. Kutsarova, G. Radonov, and R. Kalpakchieva, Izv. Akad. Nauk SSSR, Ser. Fiz. {\bf 35}, 2207 (1971), Bull. Acad. Sci. USSR, Phys. Ser. {\bf 35}, 2004 (1972)
\bibitem[Tot75]{Tot75t} K.S. Toth, W.D. Schmidt-Ott, C.R. Bingham, and M.A. Ijaz, Phys. Rev. C {\bf 12}, 533 (1975)
\bibitem[Rec85]{Rec85t} J. Recht, Y.K. Agarwal, K.P. Blume, M. Guttormsen, and H. H{\"{u}}bel, Nucl. Phys. A {\bf 440}, 366 (1985)
\bibitem[San63]{San63t} A. Santoni, A. Caruette, and J. Valentin, J. Phys. France {\bf 24}, 407 (1963)
\bibitem[Ste64]{Ste64t} F.S. Stephens, N. Lark, and R.M. Diamond, Phys. Rev. Lett. {\bf 12}, 225 (1964)
\bibitem[Ste71]{Ste71t} F.S. Stephens, J.R. Leigh, and R.M. Diamond, Nucl. Phys. {\bf A170}, 321 (1971)
\bibitem[Wil47]{Wil47t} G. Wilkinson, Phys. Rev. Lett. {\bf 160}, 864 (1947)
\bibitem[Wil50]{Wil50t} G. Wilkinson, Phys. Rev. {\bf 80}, 495 (1950)

\end{theDTbibliography}

\end{document}